\documentclass[11pt]{article}

\usepackage{amsmath}
\usepackage{amssymb}

\newcommand{\beq}{\begin{equation}}
\newcommand{\eeq}{\end{equation}}
\newcommand{\bit}{\begin{itemize}}
\newcommand{\eit}{\end{itemize}}
\newcommand{\ben}{\begin{enumerate}}
\newcommand{\een}{\end{enumerate}}
\newcommand{\la}{\langle}
\newcommand{\ra}{\rangle}

\newcommand{\ch}{{\cal{H}}}

\newcommand{\co}{{\cal{O}}}

\newcommand{\f}{\frac}

\begin{document}

\begin{titlepage}

\begin{flushright}
\today
\end{flushright}

\vspace{1in}

\begin{center}

{\bf Quantum Mechanical Observers and Time Reparametrization Symmetry}

\vspace{1in}

\normalsize

{Eiji Konishi\footnote{E-mail address: konishi.eiji.27c@st.kyoto-u.ac.jp}}

\normalsize
\vspace{.5in}

 {\em Faculty of Science, Kyoto University, Kyoto 606-8502, Japan}

\end{center}

\vspace{1in}

\baselineskip=24pt
\begin{abstract}
We propose that the degree of freedom of measurement by quantum mechanical observers originates in the Goldstone mode of the spontaneously broken time reparametrization symmetry. Based on the classification of quantum states by their non-unitary temporal behavior as seen in the measurement processes, we describe the concepts of the quantum mechanical observers via the time reparametrization symmetry.
\end{abstract}

\vspace{.7in}
 
\end{titlepage}
\section{Introduction}
This paper is based on the author's previous paper\cite{Konishi} in which the measurement processes by quantum mechanical observers were described by a model of the neural-glial networks in human brains as an observer. Here, we consider some conceptual aspects of this description. In the previous paper, we invoked the Eguchi-Kawai large $N$ reduction\cite{EK} to explain the globality of the quantum behavior of the brain by constructing a large number of gauge symmetries in the neural-glial network. There is a {{Bose-Einstein condensate}} of the evanescent photons around the perimembranous regions of neurons, which is {{crucial}} to ensure the existence of off-diagonal orders in the brain wave function. These off-diagonal orders are needed in Ricciardi-Umezawa theory\cite{RU1,RU2,RU3} and its development, as seen in the concept of the {\it{superradiative circuit}} in the brain based on the work by Jibu, Pribram and Yasue (JPY)\cite{JPY}, which is based on the papers by Fr${\ddot{{\rm{o}}}}$hlich and the early papers about the role of dipole wave quanta in living matter\cite{Froehlich1,Froehlich2,GDMV1,GDMV2,GDMV3}. We noted that the quantum system of human brain is a singular from the general view point. The quantum state in our model is compatible with the informational structure of the brain consisting of the bits of spikes due to the reduction of the classical dynamical degrees of freedom. We introduced the notion of {\it{free will}} in the state reduction. However, the positioning of observers in the quantum mechanical world and several important characteristics of consciousness were unresolved; these are explained in this paper. In the previous paper, we invoked the thesis by Penrose on state reduction\cite{Penrose}, which is a new interpretation of quantum mechanics, in the context of the neural and glial networks version of the Penrose-Hameroff scenario\cite{HP1,HP2}. In this thesis, the variance of the time increment under the effect of quantum gravity is the origin of state reduction. In this paper, as the continuance of the previous paper, we study the conceptual nature of quantum mechanical observers on the basis of this thesis.

First, we summarize the results in the previous paper.
In the Penrose thesis on the state reduction, the universal definition of a human brain-like quantum mechanical observer is found to be the triple of the nonzero quantum superposition retention time $\tau$
%(decoherence time)
 in the spatially global region due to the gauging control, the information entropy $H$ of neural-glial dynamics and the vacuum expectation values of order parameters $J$ of the ground state:
\begin{equation}(\tau, H,J)_{{{{{\cal{H}}}}}, {{V}}}\;,\label{eq:finals}
\end{equation}
for the Hamiltonian ${{\cal{H}}}$ of the system and the Hilbert space $V$ of macroscopically coherent wave functions. For example, the ability to perform quantum measurements results from the combination of $\tau$ and $H$ in the case of the learning process of $H$.
For the human brain, the elements $\tau$, $H$ and $J$ represent the primitive free will, the dominant informational brain activities (e.g., the informational processes of recognition and learning and REM (Rapid Eye Movement) sleep unlearning, that is, weakening the strengths of synaptic couplings\cite{REM1,REM2}, etc.) and the dynamical memory stores, respectively. The extension of the primitive free will is due to the Eguchi-Kawai large $N$ reduction by the action of glia cells.\cite{Konishi} The memory stores $J$, that are the quantum field theoretical synaptic couplings and reproduce the corresponding non-unitary changes, have an interrelation to the informational entropy of brain activity $H$ as seen in the Hopfield model.\cite{Hopfield} This human brain-like interrelation between $H$ and $J$ is a universal property of the definition of consciousness. In this paper, based on the criterion for human brain-like consciousness given in Eq.(\ref{eq:finals}), we investigate the conceptual aspects of quantum mechanical observers.

The construction of this paper is as follows.
In Section 2, we define the {\it{quantum class}} of quantum states by the equivalence class of their non-unitary time developments by scaling, while maintaining their variance of the time increment. We explain that the quantum mechanical world has scale and time reparametrization symmetries. We then discuss the spontaneously broken phase of time reparametrization symmetry and identify the degree of freedom of the quantum gravity effects of the time increment with the Goldstone mode of the spontaneously broken time reparametrization symmetry. In Section 3, we first introduce the notions of {\it{none}} and {\it{ourselves}} by using quantum and classical aspects of consciousness, respectively. Second, we describe the process from ourselves to none (neural death) by using the category of quantum classes. Third, we describe the transition from none to ourselves (neural birth). In the simplified demonstration of this process, we use the null-Hamiltonian constraint and invoke Vilenkin's idea of non-space-time tunneling in the quantum creation of the Universe from nothing.\cite{Vilenkin1,Vilenkin2,Vilenkin3}
 We derive a formula for the rate of this tunneling. 
 This tunneling accompanies the spontaneous breakdown of the time reparametrization symmetry, which is the arbitrariness of the parametrization of time, on the category of quantum classes. The concept of time in this process is also discussed and we remark on the sleeping state in our perspective.
\section{Quantum Classes and Reparametrization Symmetries}
Based on the Penrose thesis on state reduction\cite{Penrose}, as in our original argument, we define the {\it{quantum classes}} of the quantum mechanical wave functions by the identification of pairs of related wave functions. These wave functions are related to each other by a certain compatible complication of the renormalizations of their physical scales, that is, the space-time scale and the scales of physical quantities (e.g., the scales of {\it{time variables}} which are the coefficients in the exponential map of the conserved charge operators of the system, such as time $t$ for the Hamiltonian ${\cal{H}}$ and the rotation angle for angular momentum in a central force system, etc.). The renormalization commutes with the quantum mechanical time development as operations on the wave functions. That is, the complication of the renormalizations preserves the non-unitary effects of the quantum variance of the time increment of the wave function when its Hilbert space is transformed. Importantly, since in the Penrose thesis the stochastic variable is not the wave function but time increment, the time developments of the wave functions can transition between the different Hilbert spaces within the common quantum class without losing their physical meaning. Consequently, the wave function pairs are classified by the 
%decoherence
superposition retention time, and once we define such classes, we can distinguish the quantum mechanically trivial and nontrivial states by this 
%decoherence
superposition retention time. 
%The role of an observer is a quantum mechanically relative concept for a position in the quantum mechanical world as a quantum class.

 Here we make three observations about non-unitary processes. First, we note that discussing non-unitary processes using the notion of quantum classes is meaningful only when each quantum class being considered is related to an open quantum system. In our central arguments, we want to discuss the case where the quantum classes are the brain wave functions. Fortunately, human brains are open quantum systems and their dynamics are, in the context of the Ricciardi-Umezawa theory, characterized by dissipation. The dissipative quantum model of the human brain was theoretically founded by Vitiello and has recently received experimental support.\cite{Vitiello,FL} The whole construction of this paper is based on this openness of the brain quantum systems. Second, the non-unitary processes treated in this paper are those not in quantum mechanics but in quantum field theory; in the latter the decoherence phenomena are known to be less harmful than in the former. Third, regarding the non-unitary processes of measurement, they may be triggers of symmetry breaking in the system being considered, and place the system in a specific state space unitarily inequivalent to other state spaces.\cite{Cel,Bla}

Here, for convenience, we introduce the concept of the {{category}} of the quantum classes, denoted by $C$. In mathematics, a {\it{category}} consists of a set or class of objects and the morphisms between each pair of objects, which include the identity map for pairs of the same object and have a composition structure with associativity.\cite{Category} The category $C$ of the quantum classes is mathematically defined as follows. First, its objects are the sequences of the non-unitary temporal developments of the spaces of temporally varying quantum classes of wave functions. Second, the morphisms between objects are defined from the restrictions between the Hilbert spaces of wave functions of objects, which are unitarily inequivalent spaces. Third, the compositions of morphisms are the compositions of the transformations. From its definition, this category is time dependent.

Next, we introduce the notions of scale and time reparametrization symmetries in the quantum mechanical world. To simplify the argument, we consider integrable systems. First, the above definition of quantum classes is scale-intrinsic. Thus, in an obvious argument, it is not affected by arbitrary spatial scale reparametrizations as renormalization group-like changes of the time variables $x_a$ of conserved charge operators $Q_a$ of spatial symmetries (i.e., $d\langle Q_a\rangle/dt=0$ and $Q_a\neq {\cal{H}}$):
\begin{equation}
x_a\longrightarrow x_a^\prime\ \ {\mbox{with}}\ \ x_a\equiv f_a(x^\prime)\;,
\end{equation}
where the index $a$ runs over all of the time variables. 
Second, the wavefunction of the Universe is a solution of the Wheeler-De Witt equation\cite{WDW1,WDW2,WDW3} that is the result of the canonical quantization of gravity (to quantize we take the spatial-spatial parts of the space-time metric as variables) and matter, in the Arnowitt-Deser-Misner (ADM) decomposition of the space-time metric,\cite{ADM} applied for the null classical Hamiltonian constraint.\cite{KT} Thus, the dynamics of the wavefunction has a symmetry under arbitrary monotonic and differentiable time reparametrizations:
\begin{equation}
t\longrightarrow t^\prime\ \ {\mbox{with}}\ \ t\equiv f_t(t^\prime)\;.
\end{equation}
The scale reparametrization invariance of quantum classes and this time reparametrization symmetry mean that the quantum mechanical world does not depend on the choices for the parametrizations of time and scale variables. The time reparametrization invariance requires the absence of a Newtonian external time.\footnote{In the canonical theory of quantum gravity, since we do not separate the observing system corresponding to the coordinate frame and the observed objects, two arbitrary states linked by a diffeomorphism are equivalent to each other.}

On the other hand, when an observer uses its classical mechanical self-identity, which avoids the quantum gravity effects of time, to fix the temporal lapse function in the ADM decomposition of the space-time metric to a particular one, the time reparametrization symmetry is spontaneously broken. Indeed, even when we write down the matter Schr${\ddot{{\rm{o}}}}$dinger equations, we already assume the spontaneous breakdown of the time reparametrization symmetry (see Appendix A). In the following, we consider the spontaneously broken phase of the time reparametrization symmetry, whose existence will be shown in the next section. In this phase, a time parametrization under the spontaneously broken time reparametrization symmetry is the sum of the vacuum expectation value $\langle t\rangle$ and the Goldstone mode $\tilde{t}^G$:
 \begin{equation}t=\langle t\rangle +\tilde{t}^G\;,\ \ \langle \tilde{t}^G\rangle=0\;,\label{eq:t1}\end{equation}
where the vacuum expectation value and the Goldstone mode of time are defined by those of the temporal lapse function. (The temporal lapse function that is in the temporal-temporal part of the space-time metric\cite{ADM} is not a parameter like time but a physical quantity, and plays the role of a dynamical order parameter of the time reparametrization symmetry.)
On the other hand, the external time increment is the sum of the mean time increment $\widehat{\delta t}=\mu$ and the quantum gravitational fluctuation $\widetilde{\delta t}^Q$, treated as a normal stochastic variable:
\begin{equation}\delta t=\widehat{\delta t}+\widetilde{\delta t}^Q\;,\ \ \widehat{\widetilde{\delta t}^Q}=0\;.\label{eq:t2}\end{equation}
As will be explained in Section 3.2, $\mu$ is not constant in time when the time reparametrization symmetry is unbroken.
Then, comparing Eqs.(\ref{eq:t1}) and (\ref{eq:t2}) we propose as the main statement of this paper the equivalence
\begin{equation}\delta\tilde{t}^G=\widetilde{\delta t}^Q|_{\mu=\mu_0}\;,\label{eq:prop}\end{equation}
where we fix the mean time increment to be a constant $\mu_0$.
Namely, we state that the degree of freedom of the quantum gravity effects of the time increment, from which non-unitary temporal processes on wave functions follow in the Penrose thesis, originates in the Goldstone mode $\tilde{t}^G$ of the spontaneously broken time reparametrization symmetry. The vacuum expectation value $\langle t\rangle$ and the Goldstone mode $\tilde{t}^G$ of a time parametrization cause unitary and non-unitary time developments, respectively, in the corresponding system. As will be explained in the next section, the spontaneous breakdown of the time reparametrization symmetry is due to the fact that, though a quantum mechanical observer is described by a macroscopic quantum state, it retains the classical mechanical self-identity\cite{Konishi}, and is produced by non-space-time tunneling effects in our birth processes.

Now, we have the following perspective grounded on the above arguments:  {{The scale and time structures of the world, which we recognize by using our own scales of time variables and clocks (see Eq.(\ref{eq:unbroken2})), depend on and are formed via our own broken scale and time reparametrization symmetries.}} {{In particular, when the scale and time reparametrization symmetries of a quantum mechanical observer are broken or unbroken, the scale and time reparametrization symmetries for its perceptible surrounding world are also broken or unbroken, respectively.}} The {\it{surrounding world}} indicates its classical mechanical time development by the constant mean time increment $\widehat{\delta t}$ (see Section 3.2). So, a quantum mechanical system cannot always have its own quantum mechanical world in the above sense.

\section{Concepts of Observers}
\subsection{Self-identities in Observers}
In quantum mechanics, as each quantum with the same quantum numbers has no individuality and is a probability cloud due to the uncertainty principle, each quantum state with its $\tau$ also has no individuality (i.e., it is {\it{none}}). When we consider a set of such quanta, we cannot distinguish among them. So, to follow them in spatially different regions or to discuss where they belong does not make sense unless observers measure their positions, and there is no clear border line between them.
Their non-unitary dynamics are reduced on the fluctuation of the time increment, and via the concept of the fluctuation of the time increment, the only self-identical abstract notions are the quantum classes and their category.

We now apply this perspective to our theory. Our brains constitute macroscopically enlarged and nontrivial quantum classes. The fact that {{they have no individuality}} means that quantum recognition, which we associate with {\it{qualia}}, has a universality between brains. On the other hand, the classical mechanical notion $H$ has an individuality and is distinguishable. Namely, it depends on the classical mechanical material properties of the neural-glial network. In the Ricciardi-Umezawa theory, the memories $J$ are self-identical, in a different sense, as the vacuum expectation values of order parameters\cite{RU1}. We identify ourselves, and distinguish ourselves from others, mainly by these notions $H$ and $J$. However, we must not confuse these self-identities with the identity of the quantum class of wave functions with a particular $\tau$. Just like the relation between quantum mechanics and classical mechanics, regarding self-identity, the quantum class of wave functions with a particular $\tau$ differs essentially from these notions, and the concept of {\it{ourselves}} is an approximate classical mechanical concept.

Our consciousness as ourselves plus its none under the criterion of Eq.(\ref{eq:finals}) was derived in the birth process and accompanies the {{spontaneous breakdown of the time reparametrization symmetry}} on the category of none, that is, losing the arbitrariness in the parametrization of time. This is due to the fact that, when we a priori admit the quantum gravitational effects on the time increment (r.h.s. of Eq.(\ref{eq:prop})) and the matter Schr$\ddot{{\rm{o}}}$dinger equations for systems of observers with a formal time parameter (see Appendix A), the count $\nu$ of derived non-unitary processes with constant mean time increment $\mu_0$ (where $\nu$ and $\mu_0$ are due to the fact that our consciousness has the quantum (none) and classical mechanical (ourselves) self-identities, respectively) plays the role of a clock, which does not allow any time reparametrization and introduces the Newtonian external time. Thus, the time reparametrization symmetry is spontaneously broken (l.h.s. of Eq.(\ref{eq:prop})). This argument is compatible with Eq.(\ref{eq:prop}). We note two points. First, of course, every quantum class with a concept of itself, not only human brain-type consciousness states, has such a clock. Second, the constant time increment can be applied to general classical mechanical systems. However, in general, due to their zero 
%decoherence
quantum superposition retention time, they are trivial and out of consideration in the classification of quantum mechanical objects by quantum classes (see Section 2) and we consider only the quantum mechanical world by this classification. Furthermore, as easily noted, the birth process also accompanies the spontaneous breakdown of the scale reparametrization symmetry. These two types of broken symmetries are related to each other in Schr$\ddot{{\rm{o}}}$dinger equations but not in the Wheeler-De Witt equation.

As explained here, our conscious activities always contain an immortal, quantum mechanical and macroscopically enlarged characteristic of none in a macroscopically nontrivial quantum class of wave functions besides the living classical mechanical characteristic of ourselves, which produces this quantum class and surrounds it by a classical mechanical level potential barrier. The free will in conscious activities, defined by the fluctuating time increments, and the qualia of perceptions, defined as a high-order role of the free will in the case of the learning process of $H$, are characteristic not of ourselves but of the enlarged none. In general, the free will and the qualia survive as long as the system being considered consists of Eq.(\ref{eq:finals}), otherwise they are too faint to be detected at the classical mechanical scale, as rates of general quantum mechanical effects occur on spatial nanoscales and usually decrease exponentially, as seen in tunneling effects.

In the rest of this paper, we refer to ourselves plus the relevant none (i.e., the {\it{relic}} of pure none) as simply {\it{self}}.

\subsection{Transition Processes}
Next, we describe the transition processes between self and none (neural death or birth). First, we describe the process from self to none (neural death). When the self dies, instead of losing characteristics such as the Eguchi-Kawai large $N$ reduction in the neural-glial network\cite{Konishi}, it plays the role of a part, corresponding to a synaptic site in analogy to the neural-glial networks, in the category of none and it is a {\it{pure none state}}, which possesses the exact time reparametrization symmetry.

The time property of pure none states is characterized by
\begin{equation}\widetilde{(\mu,{\cal{T}})}_C\;,\label{eq:unbroken}\end{equation}
where ${\cal{T}}$ denotes the 
%decoherence
quantum superposition retention time of the objects in the quantum mechanical world $C$ along the ensuing world branch ruled by the Penrose thesis, and $\mu=\mu(t)$ is the mean time increment, which is not constant, due to the time reparametrization symmetry. (Here, $\mu>0$.)
Namely, the variable characterizing the time property increases from ${\cal{T}}$ to $\widetilde{(\mu,{\cal{T}})}$. Here, the tilde denotes the equivalence classification under the time reparametrization symmetry. By this increment of the number of variables, the time reparametrization symmetry is retained.

The most important point of Eq.(\ref{eq:unbroken}) is that the time reparametrization symmetry is a gauge symmetry of time. For a gauge symmetry, under its gauge equivalence, the moduli space of symmetry variables ${\cal{M}}$, which survives the gauge equivalence, contracts to ${\cal{M}}/{\cal{G}}$, where ${\cal{G}}$ is the symmetry group. In a global symmetry, its moduli space is still ${\cal{M}}$. The gauge invariant quantity of this symmetry is the count of the non-unitary changes. From these facts, in Eq.(\ref{eq:unbroken}), the unitary time evolution between two arbitrary non-unitary changes loses its quantitative sense. In other words, the pure none state is unable to recognize these unitary time developments, which are gauge equivalent to each other.

We note that the same statement holds for the relic of the pure none state (i.e., the Goldstone mode of the broken time reparametrization symmetry) that carries the core role of consciousness as already explained in detail.

Another important point of Eq.(\ref{eq:unbroken}), which was alluded to in Section 2, is that in the surrounding world of an observer, in whose wave function time reparametrization symmetry is spontaneously broken by using their own clock,  Eq.(\ref{eq:unbroken}) is a property of the quantum mechanical world itself, which contains all of the quantum mechanical phenomena along an ensuing world branch, as described by this observer:
\begin{equation}(\mu_0,{\cal{T}})_C\;,\label{eq:unbroken2}\end{equation} where the mean time increment is fixed to a constant $\mu_0$. We remark that the spatial inclusion relation between the quantum mechanical world and the observer is irrelevant to the issue of time induced by the observer's clock in Eqs. (\ref{eq:unbroken}) and (\ref{eq:unbroken2}). We recall that to describe physical phenomena by quantum mechanics, we always assume the existence of an observer's measurements. But when we describe an observer's quantum mechanical world with Eq.(\ref{eq:unbroken2}), in which time reparametrization symmetry is broken, we usually idealize the description by ignoring them. We must not confuse Eqs.(\ref{eq:unbroken}) and (\ref{eq:unbroken2}).

Next, we pursue this pure none state after neural death. As defined in the last section, the quantum classes are classified by their {{non-unitary}} behavior of the quantum time increment. There are notable features in the pure none states. First, the definition of the quantum classes is the equivalence class formed by spatial rescaling and renormalizations. Second, based on this first feature, it has the following {{plasticity}}: when the 
%decoherence 
superposition retention time of the quantum classes (i.e., the frequency of non-unitary processes) changes, this change plastically influences the category via the morphisms, due to the definition of a quantum class, to reformulate the quantum classes in order to let the non-unitary processes of a quantum class coincide with those of other quantum classes in the category $C$ (i.e., to maintain the consistency of the category $C$ as a quantum mechanical world). Due to this plasticity, the object states satisfying Eq.(\ref{eq:unbroken}) may spontaneously transit to a new temporally stable human brain-like consciousness (i.e., self) satisfying Eq.(\ref{eq:finals}) (if one exists in the environment), changing the spatial scales and time parametrization (for the pure none states, spatial scales and time parametrization are dynamical variables). This is the {{spontaneous breakdown of the scale and time reparametrization symmetries}} due to tunneling birth, as will be explained soon.

After neural death, the aspect of self is lost and only none survives. Due to Liouville's theorem, the quantum class of wave functions with the $\tau$ of this none is temporally immortal by using rescaling. Even though the self is lost, there may exist transitions from none to self. We note that the ending self and beginning self put a pure none state between them; so they do not correspond and cannot be connected. Thus, we cannot identify the self in neural birth and the self in neural death.

Finally, we describe the transition from pure none to self (neural birth).
To simplify the explanation, we consider as an example the neural-glial network\cite{Konishi} in the brain as the self. Its dynamical variables are
 the neural and glial states. The neural states are defined by the synaptic coefficients of the wave function of the globally coherent {{superradiative circuit}} in the brain and indirectly indicate spikes by their signs; a positive sign for a spike and a nonpositive sign for a non-spike,\begin{equation}
\varphi=\phi_i[k]\;,\ \ i=1,2,\ldots,n\;,\ \ k=1,2,\ldots,N\;,\label{eq:neuron}
\end{equation}
where $n$ is the number of neurons and $N$ is the span of the temporal steps counted by the time intervals of spikes.\cite{Konishi} The norm of the $n$-vector $\phi$ is normalized. The glial states ${\cal{G}}$ act on the $N$-vector neural states $\varphi$, where
\begin{equation}
{\cal{G}}\in o(N)\;.\label{eq:glia}
\end{equation}

 The birth process, that is, the process of the creation of self (including the relic of pure none) from a pure none state, is described by a non-space-time tunneling into the potential barrier of ourselves. Regarding this point, we note that the neural network, corresponding to ourselves, and the superradiative circuit, corresponding to the none, are different physical entities, though they are highly correlated with each other.\cite{Konishi} So, although this birth process is induced by classical mechanical activation on the neural network (i.e., ourselves) via ATP, glutamate and so on, the birth process for the superradiative circuit (i.e., the none) to activate it is quantum tunneling (recall the definition of the neural states $\phi$).

 To describe the activation of this none,\footnote{Here, {\it{activation}} means to coincide two asymmetries of directions of evolutions in that the system evolves as its potential energy decreases and that the system promotes its state only when neurons are fired (i.e., for positive valued $\phi$).} we use the null-Hamiltonian constraint on its wave function in the canonical quantum theory of Eq.(\ref{eq:neuron}), not in the language of operator formalism, which leads to the exact time reparametrization invariance of the initial pure none state,
 \begin{equation}(-{\hbar^2}\nabla^2_\phi+2\mu_\phi\ch_{int}(\phi))\psi(\phi)=0\;,\label{eq:Sch}\end{equation}
 where the arbitrariness of the constant additivity of the Hamiltonian 
 $\ch_{int}$ (i.e., $\ch_0$ in Eqs. (\ref{eq:H}) and (\ref{eq:H2}))
is fixed to adjust a particular non-spike state (e.g., all $\phi$ are $-\frac{1}{\sqrt{n}}$) to be a zero-energy state. 
Here, in the quantum regime, since the variables of the brain wave function are the neural states $\phi$, the glial states ${\cal{G}}$ can be recognized as a temporally constant background.
So, in Eq.(\ref{eq:Sch}), we quench the glial states ${\cal{G}}$ temporally and focus only on the neural states.
When we need to describe the activation process in a simplified situation, choosing the constraint on the none states to be the null-Hamiltonian constraint is appropriate, since the none state identified with a quantum system itself before any activation of the quantum system is the zero-energy state, as explained above. After the tunneling into activation, the quantum system temporally develops according to the rule in the Hopfield type quantum neural-glial network model\cite{Konishi} (i.e., nonlinearly rolling down the potential energy in the higher dimensional space of variables, Eqs.(\ref{eq:neuron}) and (\ref{eq:glia})).
We remark that in general exact time reparametrization invariant quantum systems, we identify the solutions of Schr${\ddot{{\rm{o}}}}$dinger equations written with equivalent time parametrizations by keeping count of non-unitary processes. When we use a general pure none state, which is assumed to be a maximally closed system, as the initial state, the process of neural birth is also described by tunneling. This is because, if it is not tunneling, the initial state is already self and contradicts the above assumption.
The definitions of the elements in the total Hamiltonian in Eq.(\ref{eq:Sch}) are as follows.
 First, $\nabla^2_\phi$ is the Laplacian.
Second, $\mu_\phi$ is the inertia of neural states measured by the kinetic responses to energy inputs. Third, when the glial states ${\cal{G}}$ are not temporally quenched,
the interaction Hamiltonian of the quantum neural-glial network, which has the {{activation}} potential barrier for none states, is\cite{Konishi} 
\begin{equation}
\ch_{int}(\varphi,{\cal{G}})=-\f{1}{2N}\la\la \varphi,\exp(\Delta)\varphi\ra\ra+\ch_0\;.\label{eq:H}\end{equation}
Here, $\la\la A,B\ra\ra$ denotes the inner product of $A_i[k]$ and $B_i[k]$ formed by contracting on both $i=1,2,\ldots,n$ and $k=1,2,\ldots,N$, and 
 \begin{equation}
 \Delta \varphi=\delta\varphi+{\cal{G}}\varphi\;,
 \end{equation}
 where $\delta$ denotes the difference regarding the index of a neuron's sites.
 Eq.(\ref{eq:H}) has an $O(N)$ gauge symmetry for the local index $i$ of the transformations:
\begin{equation}
\varphi\to\co_i \varphi\;,\ \ {\cal{G}}\to \co_i {\cal{G}}\co_i^{-1}-(\delta \co_i)\co_i^{-1}\;,\ \ \co_i\in O(N)\;.\end{equation}
Note that
 when we assume $N$ is large, the Eguchi-Kawai large $N$ reduction
\begin{equation}
\exp(\Delta)\to\exp({\cal{G}})\;,
\end{equation}
can be applied to the partition function of the Hamiltonian ${\cal{H}}_{int}$.
When the glial states ${\cal{G}}$ are temporally quenched, the interaction Hamiltonian $\ch_{int}$ can be written as
\begin{equation}
\ch_{int}(\phi)=-\frac{1}{2}\la \phi,J\phi\ra+\ch_0\;,\label{eq:H2}
\end{equation}
%changed
where the {{effective}} synaptic couplings are denoted by $J$ and assumed to be temporally quenched with respect to the quantum regime.
Here, $\la A,B\ra$ denotes the inner product of $A_i$ and $B_i$ formed by contracting on $i=1,2,\ldots,n$.

In the activation process of the none, the wave function begins from a particular pure none state $\psi(\phi_0)$ with negative-valued $\phi_0$, just like Vilenkin's scenario of the birth of the Universe from nothing. This kind of scenario leads to the absence of an initial singularity.  
 In $\psi$, $\phi$ is not a function but just a number. So, the potential is not a function of space-time coordinates $(x,t)$, reflecting the scale and time reparametrization invariance of the initial pure none state. This is the meaning of {{non-space-time}} in `non-space-time tunneling'.
 Now, using Eq.(\ref{eq:Sch}), we derive the formula for the activation rate (i.e., the tunneling rate) for the potential barrier $\ch_{int}$ in the following simplified situation.
When we discuss the tunneling process of the brain wave function, in terms of the behavior of the neural states, the radial part is dominant.\footnote{Here, we relax the normalization condition on the vector $\phi$. After we obtain the vector $\phi=\langle\phi\rangle$ by using $\psi(\phi)$, we normalize it so we can compare the results of $\psi(\phi)$ with those of the wave function $|0(\beta)\rangle$ defined in Ref.1.} So, to simplify the argument, we consider the case in which the system depends only on the radius of the space of the neural states, denoted by $\phi_r$, and the system is reduced to a one-dimensional one. We denote the strength of the reduced effective synaptic couplings by $J_0$. Then, Eq.(\ref{eq:Sch}) is reduced to
 \begin{equation}
 (-{\hbar^2}\nabla_{\phi_r}^2+2\mu_\phi\ch_{int}(\phi_r))\psi(\phi_r)=0\;,\ \ \ch_{int}(\phi_r)=-J_0\phi_r^2+\ch_0\;.\label{eq:Sch4}
 \end{equation}
 When we apply the WKB approximation to the tunneling process for the potential $\ch_{int}$ in Eq.(\ref{eq:Sch4}), the well-known results in the general setting for the in-coming wave function $\psi_{in}$ from the state $(\phi_0)_r$, the transmitting wave function $\psi_{tr}$ and the out-going wave function $\psi_{out}$ are\cite{Landau}
 \begin{subequations}
 \begin{align}
 \psi_{in}(\phi_r)&=e^\Lambda\frac{(-ic)}{\sqrt{p(\phi_r)}}\exp\biggl[i\biggl(\frac{1}{\hbar}\int_{\phi_r}^{a}p(\phi^\prime_r)d\phi^\prime_r-\frac{\pi}{4}\biggr)\biggr]\;,\\
 \psi_{tr}(\phi_r)&=\frac{-ic}{\sqrt{\varrho(\phi_r)}}\exp\biggl(-\frac{1}{\hbar}\int_{b}^{\phi_r}\varrho(\phi_r^\prime)d\phi_r^\prime\biggr)\;,\\
 \psi_{out}(\phi_r)&=\frac{c}{\sqrt{p(\phi_r)}}\exp\biggl[i\biggl(\frac{1}{\hbar}\int_{b}^{\phi_r} p(\phi_r^\prime)d\phi_r^\prime-\frac{\pi}{4}\biggr)\biggr]\;,\label{eq:out}
 \end{align}
 \end{subequations}for $a=-\sqrt{\ch_0/J_0}$, $b=\sqrt{\ch_0/J_0}$, a complex number $c$, and
 \begin{subequations}
 \begin{align}
 p(\phi_r)={\textstyle{\sqrt{-2\mu_\phi\ch_{int}(\phi_r)}}}\;,\ \ \varrho(\phi_r)={\textstyle{\sqrt{2\mu_\phi \ch_{int}(\phi_r)}}}\;,\ \ \Lambda=\frac{1}{\hbar}\int_{a}^{b}\varrho(\phi_r^\prime)d\phi_r^\prime\;.
 \end{align}
 \end{subequations}
 Then, for the currents of probability densities $j_{in}$ and $j_{out}$, the formula for the activation rate $T$ is
 \begin{eqnarray}
 T&=&\frac{j_{out}}{j_{in}}=e^{-2\Lambda}\nonumber\\
 &=&\exp\biggl(-\frac{2}{\hbar}\int_{a}^{b}{\textstyle{\sqrt{2\mu_\phi(-J_0(\phi^\prime_r)^2+\ch_0)}}}d\phi^\prime_r\biggr)\nonumber\\
 &=&\exp\biggl(-\frac{\ch_0}{\hbar}\sqrt{\frac{8\mu_\phi}{J_0}}\int_{-1}^1\textstyle{\sqrt{1-(\phi^\prime_r)^2}}d\phi_r^\prime\biggr)\nonumber\\
 &=&\exp\biggl(-\frac{\pi\ch_0}{\hbar}\sqrt{\frac{2\mu_\phi}{J_0}}\biggr)\;.\label{eq:tr}
 \end{eqnarray}
 (However, this is just a theoretical result, and to obtain the numerical value of $T$ we need experimental measurements of three parameters for the human brain: $\mu_\phi$, $J_0$ and $\ch_0$.)
Here we make two remarks. First, the idea of non-space-time tunneling can be also applied to neural death when the dying process is governed by Eq.(\ref{eq:Sch}) by reversing the positions and the directions of the in-coming and out-going wave functions in neural birth. Since the neural-glial system promotes its state only when neurons are fired, the tunneling of neural death cannot happen spontaneously.
Second, after the activation, the mechanism of the Eguchi-Kawai large $N$ reduction and the presence of a Bose-Einstein condensate, which ensures the existence of the off-diagonal orders, are the necessary conditions for the new-born human brain-like consciousness to satisfy the criterion in Eq.(\ref{eq:finals}). On this point, we note that the quantum mechanical properties of conscious activities, such as the dynamics of their quantum classes, {{all depend}} on whether there is a Bose-Einstein condensate. Actually, based on the JPY model, it has been hypothesized by Jibu that, during anesthesia, the order structure of water molecules in the perimembranous regions of neural cells is broken by the anesthesia molecules and the critical temperature of the Bose-Einstein condensate falls to less than the living body temperature.\cite{Jibu} When we are using the standpoint of JPY model, where the Bose-Einstein condensate is considered to be directly connected with the physical substance of consciousness, this hypothesis explains the pressure reversal of the potency of anesthesia molecules\cite{Rev}. The experience of anesthesia mediates the pure none. Only in this case the selves mediating a pure none correspond.

Here we consider a time paradox. As alluded to before, in the category of pure none states there is no Newtonian external time as is in the Wheeler-De Witt equation\cite{KT}. In the above scenario of neural birth, the brain wave functions $\psi(\phi)$ correspond to this situation. This paradox is resolved for quantum mechanical observers, as already mentioned, when we admit the matter Schr$\ddot{{\rm{o}}}$dinger equations with a formal time parameter, by adopting the counting of non-unitary processes as a clock after the spontaneous breakdown of the time reparametrization symmetry. The matter Schr$\ddot{{\rm{o}}}$dinger equations are derived from the Wheeler-De Witt equation in the semiclassical regime of the expanding Universe by adopting the growing scale factor of the Universe as a formal time parameter (see Appendix A).\cite{Vilenkin4}

From our perspective, we make a remark about the sleeping brain state.
The spontaneous breakdown of time reparametrization symmetry is accompanied by non-unitary processes within a self, which has a fixed mean time increment, that is, {{measurements}}. The absence of measurement recovers the broken time reparametrization symmetry. Since sleeping state is considered to have no measurement process, due to the blocking of sensory inputs, that is, real world data\cite{Dream}, except for reproduction from the memory stores, in it the time reparametrization symmetry may be recovered partially. Thus, under the unlearning activities of REM sleep, viewed as internal random inputs to forget and stabilize the memories $J$\cite{REM1,REM2}, the time concept in a sleep state may be recognized as that in a pure none state satisfying Eq.(\ref{eq:unbroken}) partially.

\section{Summary}
In any quantum mechanical observer, there are two levels of consciousness. 
We call them ourselves and the {{relic}} of a pure none state (i.e., the Goldstone mode of the spontaneously broken time reparametrization symmetry around the vacuum expectation valued time parametrization, where the spontaneous breakdown of the time reparametrization symmetry of the pure none state is produced by the {{birth}} process). The former corresponds to the vacuum expectation valued time increment and its causal dependence on the history of time is unitary, and the latter corresponds to the Goldstone mode of the time increment and its causal dependence on the history of time is non-unitary.
After neural death,
the quantum state of the observer becomes
 pure none and has no individuality. Since the time reparametrization symmetry is a gauge symmetry of time, the pure none state with exact time reparametrization symmetry and the relic of it are unable to distinguish unitary time processes between two arbitrary non-unitary changes. We classified the none, that is, the wave functions by their non-unitary temporal behavior for the quantum variance of the time increment under the equivalence produced by spatial rescaling of the spatial time variables and renormalizations of the wave functions. The birth of self is a non-space-time tunneling into an activation potential  barrier.
 This tunneling leads to the absence of an initial singularity, and the idea of  non-space-time tunneling can be also applied to neural death.
  This birth places constraints on the pure none state as the result of the spontaneous breakdown of the time reparametrization symmetry. This is due to the fact that counting non-unitary processes with a classical mechanical constant time increment plays the role of a clock, which does not allow any time reparametrization. The versions of ourselves existing before neural death and after neural birth are unrelated to each other.
 Since the time reparametrization symmetry is broken by measurement processes, the time concept in a sleep state is recognized as that in a pure none state, partially.

\begin{appendix}
\section{Brief Account for the Wheeler De-Witt Equation}
This paper is written primarily for condensed matter physicists. So, in this appendix, we present a brief account of the basic ideas of the Wheeler-De Witt equation, which is purely a concept in quantum gravity research.

The {\it{Wheeler-De Witt equation}} of the wave functions of the Universe $\Psi$\cite{WDW1,WDW2,WDW3}, which is the temporal part of the 
diffeomorphism invariance (i.e., the time reparametrization invariance) 
condition on $\Psi$, takes the form
\begin{equation}{\cal{H}}\Psi(h,{\mbox{Other\ Variables}})=0\;,\end{equation}
for the quantum Hamiltonian operator ${\cal{H}}$ of general relativity. The quantum Hamiltonian operator is obtained by the canonical quantization of general relativity about the spatial metric $h$ under the ADM decomposition of the space-time metric\cite{ADM}, which treats the temporal development of space-time as foliations of three-dimensional hypersurface by introducing two kinds of Lagrange multipliers in the Einstein-Hilbert Lagrangian, that is, a temporal lapse function and a spatial shift vector in the space-time metric. To solve the Wheeler-De Witt equation, in the full infinite dimensional moduli space (called {\it{superspace}}) of all variables in the spatial metric is not possible. So, usually, we consider the mini-superspace of the scale factor variable $a$ in the spatial metric $h$ in the context of a homogeneous and isotropic Universe.
When we write down the ``kinetic'' part of the Wheeler-De Witt equation, there is an operator ordering issue for the scale factor $a$. By choosing a certain operator ordering for the scale factor $a$, the Wheeler-De Witt equation is
\begin{equation}(-\hbar^2\nabla_a^2+U(a)+{\cal{H}}_q)\Psi(a,q)=0\;,\end{equation}
where the ``potential'' part in the absence of a cosmological term is
\begin{equation}U(a)=-\sqrt{h(a)}R^{(3)}(a)\;,\end{equation}
for the spatial scalar curvature $R^{(3)}$, and $q$ and ${\cal{H}}_q$ are the quantum mechanical variables of the matter systems being considered and their Hamiltonian, respectively.

In the semiclassical regime for the variable $a$, the scale factor of the Universe is treated as a clock. We now explain the derivation of the Schr${\ddot{{\rm{o}}}}$dinger equation for the matter wave function $\chi$, in this semiclassical regime for the scale factor only, following Vilenkin's paper.\cite{Vilenkin4}
We adopt the WKB form ansatz for the wave function of the Universe:
\begin{equation}\Psi(a,q)=A(a)e^{iS(a)/\hbar}\chi(a,q)=\psi_0(a)\chi(a,q)\;.\label{eq:ansatz}\end{equation}
Here $S(a)$ is of order $\hbar^0$. For the $\psi_0$ part of the ansatz in Eq.(\ref{eq:ansatz}), the Wheeler-De Witt equation gives
\begin{equation}(-\hbar^2 \nabla_a^2+U(a))\psi_0(a)=0\;.\end{equation}
To order $\hbar^0$ it is
\begin{equation}-(\nabla_a{S})^2+U(a)=0\;,\end{equation}
and the part of order $\hbar^1$ is
\begin{equation}iA\nabla_a^2{S}+2i\nabla_aA\nabla_a S =0\;.\end{equation}
The matter part of the Wheeler-De Witt equation is
\begin{equation}2i\hbar\nabla_aS\nabla_a\chi-{\cal{H}}_q\chi=0\;.\label{eq:Sch1}\end{equation}
By recognizing the expanding Universe as a clock, we introduce time $t$ by
\begin{equation}i\hbar\partial_t a=2N\nabla_a S\;,\label{eq:Sch2}\end{equation}
for lapse function $N=N(t)$ that is the Lagrange multiplier representing the arbitrariness of the time coordinate, and due to the time reparametrization invariance, $dt$ can appear only in the combination $N(t)dt$. When we consider $t$ as the cosmic time, $N=1$. Here we note that, in Eq.(\ref{eq:Sch2}), $\nabla_aS$ is recognized as a canonically dual variable of the scale factor in the mini-superspace.
By substituting Eq.(\ref{eq:Sch2}) in Eq.(\ref{eq:Sch1}), we obtain
\begin{equation}i\hbar \partial_t\chi=N{\cal{H}}_q\chi\;.\label{eq:Sch3}\end{equation}
When the time reparametrization symmetry is broken and we choose $N=1$, Eq.(\ref{eq:Sch3}) is the Schr${\ddot{{\rm{o}}}}$dinger equation for the matter wave function $\chi$.

\end{appendix}

\end{document}